
\input phyzzx
\catcode`\@=11 
\def\NEWrefmark#1{\step@ver{{\;#1}}}
\catcode`\@=12 
%

\def\square{\kern1pt\vbox{\hrule height 1.2pt\hbox{\vrule width 1.2pt\hskip 3pt
   \vbox{\vskip 6pt}\hskip 3pt\vrule width 0.6pt}\hrule height 0.6pt}\kern1pt}

\def\bra#1{\langle #1 |}
\def\ket#1{| #1 \rangle}
\def\vev#1{\langle #1 \rangle}
\def\for{{\rm for}}

\def\s{{\cal S}}

\def\p{\partial}

\def\wh{\widehat}

\def\s{{\cal S}}
\def\p{\partial}
\def\wp{\p^{\,r}}
\overfullrule=0pt
\baselineskip 13pt plus 1pt minus 1pt
\nopubblock
{}~ \hfill \vbox{\hbox{MIT-CTP-2240}\hbox{TIFR-TH-93-38}
\hbox{hep-th/9309027} \hbox{August 1993}}\break
\titlepage
\title{A NOTE ON GAUGE TRANSFORMATIONS IN }
\titlestyle{BATALIN-VILKOVISKY THEORY}
\author{Ashoke Sen \foot{E-mail  address: sen@tifrvax.tifr.res.in,
sen@tifrvax.bitnet } }
\address{Tata Institute of Fundamental Research \break
Homi Bhabha Road, Bombay 400005, India}
\andauthor
{Barton Zwiebach \foot{E-mail address: zwiebach@irene.mit.edu,
zwiebach@mitlns.bitnet.\hfill\break Supported in part by D.O.E.
contract DE-AC02-76ER03069.}}
\address{Center for Theoretical Physics \break
LNS and Department of Physics\break
MIT, Cambridge, MA 02139, U.S.A.}

\abstract{We give a generally covariant description, in the sense
of symplectic geometry, of gauge transformations in Batalin-Vilkovisky
quantization. Gauge transformations exist not only at the classical
level, but also at the quantum level, where they leave the action-weighted
measure $d\mu_S \equiv d\mu e^{2S/\hbar}$ invariant. The quantum gauge
transformations and their Lie algebra are $\hbar$-deformations of the
classical gauge transformation and their Lie algebra.
The corresponding Lie brackets $[\,\,,\,]^q$, and $[\,\,,\,]^c$,
are constructed in terms of the symplectic structure and the measure
$d\mu_S$. We discuss closed string field theory as an application.}

\endpage


\def\define#1#2\par{\def#1{\Ref#1{#2}\edef#1{\noexpand\refmark{#1}}}}
\def\con#1#2\noc{\let\?=\Ref\let\<=\refmark\let\Ref=\REFS
         \let\refmark=\undefined#1\let\Ref=\REFSCON#2
         \let\Ref=\?\let\refmark=\<\refsend}

\let\refmark=\NEWrefmark

\define\zwiebachlong{B. Zwiebach, `Closed string field fheory: Quantum
action and the Batalin-Vilkovisky master equation', Nucl. Phys. {\bf B390}
(1993) 33, hep-th/9206084.}

\define\wittenab{E. Witten, `A note on the antibracket formalism', Mod.
Phys. Lett. {\bf A5} (1990) 487.}

\define\ghoshalsengauge{D. Ghoshal and A. Sen, `Gauge and general coordinate
invariance in non-polynomial closed string field theory', Nucl. Phys.
{\bf B380} (1992) 103.}

\define\hatazwiebach{H. Hata and B. Zwiebach, `Developing the covariant
Batalin-Vilkovisky approach to string theory', MIT-CTP-2184, to appear in
Annals of Physics, hep-th/9301097.}

\define\witten{E. Witten, `On background independent open-string field theory',
Phys. Rev. {\bf D46} (1992) 5467, hep-th/9208027.}

\define\schwarz{A. Schwarz, `Geometry of Batalin-Vilkovisky quantization', UC
Davis preprint, hep-th/9205088, July 1992. }

\define\getzler{E. Getzler, `Batalin-Vilkovisky algebras and two-dimensional
topological field theories', MIT Math preprint, hep-th/9212043.}

\define\wittenzwiebach{E. Witten and B. Zwiebach, `Algebraic structures
and differential geometry in 2D string theory',
Nucl. Phys. {\bf B377} (1992) 55. hep-th/9201056.\hfill\break
E. Verlinde, `The master equation of 2D string theory'
IASSNS-HEP-92/5, to appear in Nucl. Phys. B. hep-th/9202021.\hfill\break
T. Kimura, J. Stasheff and A. Voronov, `On operad structures of moduli
spaces and string theory', hep-th/9307114.}

\define\senzwiebach{A. Sen and B. Zwiebach, `Local background independence
of classical closed string field theory', MIT preprint, CTP\#2222,
submitted to Nucl. Phys. B,  hep-th/9307088.}

\define\senzwiebachtwo{A. Sen and B. Zwiebach, `Quantum background independence
of closed string field theory', MIT preprint, CTP\#2244, to appear.}

\define\batalinvilkovisky{I. A. Batalin and G. A. Vilkovisky, Phys. Rev.
{\bf D28} (1983) 2567.}

\define\henneauxteitelboim{M. Henneaux and C. Teitelboim, `Quantization of
Gauge Systems', Princeton University Press, Princeton, New Jersey, 1992.}

\subsection{Introduction}
In the antibracket, or Batalin-Vilkovisky (BV) formalism, the master action
has long been known to determine not only the BRST
transformations but also the gauge transformations. Indeed, as explained
in the original paper of Batalin and Vilkovisky
[\batalinvilkovisky], and elaborated upon in a recent monograph
(Ref.[\henneauxteitelboim], \S17.4.2), the gauge transformations
of a field  (or antifield) $\Phi^i$ are
$$\delta \Phi^i = \bigl(\, \omega^{ij}\p_j\wp_k S \,\bigr)
\,\Lambda^k \, ,\eqn\one$$
where $\omega$ is the symplectic form, $S$ is the classical
master action, and $\Lambda^k$ are field/antifield independent parameters
of local gauge transformations with statistics $(-)^{k+1}$.
\foot{$\p_j$ and $\wp_j$ stand for ${\p^l \over \p\Phi^j}$ and
${\p^r\over \p\Phi^j}$ respectively, where the supercripts $l$ and $r$ denote
left and right derivatives.}
This result, however, is not completely general.
In addition to leaving only the classical master action invariant,
the above formula is not covariant
under a change of basis; it requires the use of Darboux
coordinates which make the components $\omega^{ij}$ of the symplectic
form constant. More seriously, when
$\Lambda^k$ is field/antifield dependent, the above transformations
do not generally leave the symplectic form invariant, and therefore
they do not qualify as true gauge transformations or true invariances
(we recall that in BV quantization the physics
is determined by the action {\it and} the symplectic structure).
This is in contrast to ordinary gauge theory where gauge parameters can
be chosen to be field dependent, in addition to being spacetime dependent.
If the $\Lambda^k$'s are field/antifield independent we find
$$\delta \Phi^i =  \omega^{ij}\p_j\Bigl(\,\bigl( \wp_k S \bigr)
\Lambda^k \Bigr)
=\, \bigl\{\, \Phi^i \, , \, (\wp_k S )\,\Lambda^k \, \bigr\} \eqn\two$$
showing that the gauge transformations are canonical transformations,
and are generated by the hamiltonian $K=(\wp_k S )\,\Lambda^k$. This
form of the gauge transformations has been widely used since
most field theories have been formulated using Darboux
coordinates.

The antibracket formalism has been recently formulated covariantly in the
sense of symplectic geometry. While such covariant description was known
for the classical part of the formalism, the quantum part required the
introduction of extra geometrical structure [\schwarz ].
In the covariant formalism we need not use
Darboux coordinates, and should be able to
give the form of the gauge transformations when the $\omega^{ij}$'s are
not constants. The relevant formula was given in [\witten]
$$\delta' \Phi^i = \bigl( \omega^{ij}\p_j\wp_k S \bigr) \,\Lambda^k
\, + \, \half\bigl( \wp_k \omega^{ij}\bigr) \,\Lambda^k
\wp_j S \, ,\eqn\threea$$
giving a variation of the form
$\delta' S \, =\, \half\,(\wp_k \, \{ S , S \}) \,\Lambda^k\,$,
which vanishes on account of the classical master equation.
While \threea\ gives an invariance of the classical master
action, it does not necessarily leave the antibracket invariant.
We do not have, therefore, a parametrization of the allowed gauge
transformations. Moreover, with
second order partial derivatives, and
partial derivatives of the components of the symplectic form,
Eqn.\threea\ is noncovariant.

In this paper we shall (i) write down the classical gauge transformations
and their Lie algebra (with the associated Lie bracket $[\,\,,\,]^c$)
in the general case when the gauge transformation parameters
may be field dependent, and, (ii) generalize this to the full quantum theory,
where we find a Lie algebra of quantum gauge transformations
(with the associated Lie bracket $[\,\,,\,]^q$).
The proper geometrical interpretation of the gauge parameters $\Lambda^k$
is seen to be that of hamiltonian vectors arising from some hamiltonian
$\Lambda$. In our picture, the gauge parameters are taken to be the
hamiltonians,
which are necessarily field/antifield dependent functions.
Standard gauge transformations arise from hamiltonians linear in the
fields (or antifields); by including all possible field/antifield
dependent hamiltonians we
are naturally led to Lie algebras.
The construction is
fully covariant and involves only the
symplectic structure and the action weighted measure $d\mu_S\equiv
d\mu e^{2S/\hbar}$, left invariant by the quantum
gauge transformations.
We were led to consider this measure and the notion of quantum
gauge transformations in our study of background independence of quantum
closed string field theory [\senzwiebachtwo].

\subsection{A path integral measure}
In the covariant description of the antibracket formalism a measure
$d\mu$ in the space of field/antifield configurations is necessary.
This measure is used to define the operator $\Delta_{d\mu}$.
If $d\mu = f(\Phi)\prod_i d\Phi^i$, then
$ \Delta_{d\mu} A \equiv {1\over 2f} (-1)^i \p_i(f\omega^{ij}\p_j A)\,$.
For any {\it arbitrary} measure
$d\mu$ the following identities hold [\schwarz,\wittenab,\hatazwiebach]
$$\Delta_{\rho d\mu} A = \Delta_{d\mu}\, A
+ \half \, \{ \, \ln\rho\, , A \} \ .
\eqn\chmeas$$
$$\Delta_{ d\mu}\{A,B\} = \{\Delta_{ d\mu} A, B\}
+ (-)^{A+1}\{A, \Delta_{ d\mu} B\}\ .\eqn\leibnitz$$
Other identities we will use are the exchange property
$\{A , B\}=(-)^{AB+A+B} \{B, A\}$,
and the Jacobi identity
$ (-)^{(A+1)(C+1)}\{ A , \{B , C\}\}+{\rm cyclic} =0.$

A volume element $d\mu$ is consistent if $\Delta_{d\mu}^2 =0$.
Assume we have a consistent volume element $d\mu$, and
consider the measure $d\mu_S \, \equiv \, d\mu\,e^{2S/\hbar}$.
The associated delta operator, making use of \chmeas, is found to be
$$\Delta_{d\mu_S} = \Delta_{d\mu} +
{1\over \hbar}\,\{ S\, ,\,\, \cdot \,\, \}\,. \eqn\three$$
Following [\schwarz] one can then show that
$\Delta_{d\mu_S}^2  = {1\over \hbar^2}\,
\bigl\{ \, \hbar\,\Delta_{d\mu} \,S\,+ {1\over 2}\{\, S\,,\,S\,\}\,,\,\,
\cdot \,\, \bigr\}\,,$
which indicates that $\Delta_{d\mu_S}^2$ is a linear operator,
in fact, a hamiltonian vector.
This equation also shows that $d\mu_S$ is a consistent measure
($\Delta_{d\mu_S}^2 =0$) if $d\mu$ is consistent, and, in addition,
$S$ satisfies the quantum master equation:
$\half\{S,S\}+\hbar\Delta_{d\mu}S=0$.
The operator $\Delta_{d\mu_S}$ coincides with the operator $\sigma$
discussed in [\henneauxteitelboim], here we have only pointed out its
geometrical interpretation as the delta operator of the particularly
relevant measure $d\mu_S$. The fact that the master equation can be
encoded in the consistency condition for a measure was noticed in [\getzler]
and is implicit in [\schwarz].
We observe now that the measure $d\mu_S$ is a rather fundamental object
in the BV formalism. There is no
need that this measure should be written in terms of another consistent
measure $d\mu$ and a nontrivial function $S$ satisfying the quantum
master equation. To argue this we must look at the definition of observables.

The observables in a theory are defined by
$\vev{A}\equiv \int_L \, d\lambda_S \, A$,
where $L$ denotes a Lagrangian submanifold,
defined by the condition that at any point $p\in L$, for any two
tangent vectors $e_i, e_j \in T_pL$, we have $\omega (e_i , e_j ) = 0$.
The measure $d\lambda_S \equiv  d\lambda e^{S/\hbar}$, can be defined directly
in terms of $d\mu_S$ using the same prescription  that gives us
$d\lambda$ in terms of $d\mu$ [\schwarz]. Let $p\in L$, and $(e_1,\ldots,e_n)$
be a basis of $T_pL$. One then defines
$$d\lambda_S (e_1,\cdots,e_n)\equiv [d\mu_S (e_1,\cdots,e_n,f^1,
\cdots,f^n)]^{1/2}\ ,\eqn\LAMBDA$$
where the vectors $f^i$ are any set of tangent vectors of the full manifold
at $p$
satisfying $\omega(e_i,f^j)=\delta_i^j$. This condition fixes the vectors
$f^j$ up to the transformation $f^j \to f^j + C^{ji} e_i$. The right hand
side of \LAMBDA, however, is invariant under this transformation, since
it corresponds to a transformation of the complete basis
$(\{ e_i\} ; \{ f^j\})$ by a matrix of unit superdeterminant.
Eqn.\LAMBDA\ gives us the path integral measure of the gauge fixed theory
in terms of $\omega$ and $d\mu_S$.
Finally,  in order for $\vev{A}$ to be
independent of the choice of lagrangian submanifold,
$A$ must be a function of fields/antifields satisfying
$ \hbar\, \Delta_{d\mu} A + \{ S , A\} = 0\,\to\,\Delta_{d\mu_S}\, A = 0$
(by Eqn.\three).
This condition defines physical operators in terms of
$d\mu_S$ and $\omega$.

\subsection{Classical gauge transformations}
If the gauge transformations are to be symplectic they must be generated
by a hamiltonian function. We should therefore have
$$\delta \Phi^i = \, \{ \Phi^i \, , K \,\} \,, \eqn\four$$
for some suitable {\it odd} function $K$.
The symmetries of the BV action are generated by $K$'s for which
$\delta S = \{ S \, , K \, \}=0$. Since this is the condition for $K$ to be
a classical observable, to every local observable,
we can associate
a local gauge symmetry of the classical theory. A special class of
solutions are given by the trivial observables
$$K = \{ S\, , \Lambda\}\, ,\eqn\seven$$
since $\{S  , K \}  =0$,
by virtue of the Jacobi identity and the classical master equation.
Here $\Lambda$ is an even function of $\Phi^i$. As we will see, standard
gauge transformations arise from trivial observables $K$.
We will discuss later, in the context of string theory, why nontrivial
observables do not lead generically to standard gauge transformations.

Let us now see how we recover the gauge transformations
in Eqn.\one. In a Darboux frame,
$\Lambda = \Phi^i \omega_{ik} \Lambda^k$, with $\Lambda^k$
constants, leads to  $K= (\wp_k S)\Lambda^k$. Back in
Eqn.\four\ we then recover the original form of the gauge transformations.
Note that,
by construction, $\Lambda^k$ is the hamiltonian vector associated
to the hamiltonian $\Lambda$. We have therefore found that the original
parameters of gauge transformations have the geometrical interpretation
of hamiltonian vectors.
It is also instructive to understand the way the
transformations given in \threea\ fit into this description.
Such transformations are easily rewritten as
$$\delta' \Phi^i = \bigl\{ \Phi^i \, , (\wp_k S ) \Lambda^k \bigr\}
\, - \omega^{ij} ( \p_j \Lambda^k ) (\wp_k S)\, + \, \half\bigl(
\wp_k \omega^{ij}\bigr) \,\Lambda^k
\wp_j S \, ,\eqn\threeb$$
We now recall that a transformation of the type
$\delta_t \Phi^i = (\wp_j S)\, \mu^{ij}$ leaves $S$ invariant in a trivial
fashion if $\mu^{ij} = (-)^{ij+1} \mu^{ji}$, although it does not
necessarily leave $\omega$ invariant.
One can verify that the second and third term in
the above equation {\it do not} correspond in general
to trivial transformations. However, if $\Lambda^k$ is hamiltonian
($\Lambda^k = \omega^{kj}\p_j \Lambda$) they do. In this case we can prove that
\foot{We use the Jacobi identity
as well as the identities in Appendix B of [\hatazwiebach].}
$\delta' \Phi^i = \bigl\{ \Phi^i \, , (\wp_k S ) \Lambda^k \bigr\}
\,+ (\wp_j S)\, \mu^{ij}\,,$
with $\mu^{ij} = \half (-)^i \bigl[ \omega^{jk}\p_k\Lambda^i + (-)^{(i+1)(j+1)}
\omega^{ik}\p_k\Lambda^j\bigr]$. Therefore, when $\Lambda^k$ is
hamiltonian the $\delta'$ transformations
differ from gauge transformations by trivial transformations.
Note, however, that the $\delta'$ transformation still cannot be regarded
as a genuine symmetry of the BV theory since it does not represent in general
a canonical transformation.

Let us now compute the algebra of gauge transformations.
Let $\wh\delta_K$ denote the canonical transformation generated by the
odd function
$K$: $\wh\delta_K f=\{f, K\}$. Given that
$[\wh\delta_{K_2}\, , \, \wh\delta_{K_1}] =
\wh\delta_{\{K_1,K_2\}}$,  canonical transformations generated by observables
form a Lie algebra, since we have
$\{S, \{K_1, K_2\}\}=0$, whenever
$K_1$ and $K_2$ are
observables. While any local observable
in the theory generates a symmetry, not much can be said about the
algebra of non-trivial observables unless we know which specific theory
we are studying. Hence we shall focus our attention on the Lie subalgebra
whose elements are the canonical transformations
generated by trivial observables $K=\{S, \Lambda\}$.
Let $\delta_{\Lambda}^c = \wh\delta_{\{ S, \Lambda \}} $ denote
the classical gauge transformations induced
by $\Lambda$, then
$\bigl[\, \delta_{\Lambda_2}^c \, , \, \delta_{\Lambda_1}^c\, \bigr]
= \bigl[\, \wh\delta_{\{ S, \Lambda_2 \}} \, , \,
\wh\delta_{\{ S, \Lambda_1 \}}\, \bigr] = \wh\delta_{\,\bigl\{ \,
\{ S, \Lambda_1 \}\,,\,\{ S, \Lambda_2 \} \,\bigr\}}$.
Moreover, the Jacobi identity and the master equation imply that
$\bigl\{ \,\{ S, \Lambda_1 \}\,,\,\{ S, \Lambda_2 \} \,\bigr\} =
\half \bigl\{ \, S,\, \bigl\{ \Lambda_1 ,\,\{ S, \Lambda_2 \}\bigr\} \,\bigr\}
\,-\,\half \bigl\{ \, S,\, \bigl\{ \Lambda_2 ,\,\{ S, \Lambda_1 \}\bigr\}
 \,\bigr\}\,$. As a consequence we have that
$$\bigl[\, \delta_{\Lambda_2}^c \, , \,
 \delta_{\Lambda_1}^c\, \bigr]
= \delta^c_{[\Lambda_1,\Lambda_2 ]^c} \,, \quad
[\Lambda_1,\Lambda_2 ]^c \equiv \half \bigl\{ \, \Lambda_1 ,\,
\{ S, \Lambda_2 \} \,\bigr\}
-\half\bigl\{ \, \Lambda_2 ,\,\{ S, \Lambda_1 \} \,\bigr\} \,.\eqn\tent$$
In contrast with the standard description
of gauge transformations via generating sets of transformations
which do not give a Lie algebra, and close only up to trivial symmetries
(see [\henneauxteitelboim]), our description of gauge
transformations arising from hamiltonian functions $\Lambda$ gives directly
a closed Lie algebra. The commutator of two gauge transformations is
a gauge transformation of the same type.
The usual gauge transformations, arising from $\Lambda$'s that are
only linear in fields or antifields, as expected, do not close among
themselves in general.
They close when supplemented
with $\Lambda$'s having additional field/antifield dependence.
 The trivial identity
$\bigl[\, [ \delta_{\Lambda_3}^c \, , \,\delta_{\Lambda_2}^c\,] ,
\delta_{\Lambda_1}^c \bigr] + \hbox{cyclic} = 0$, implies that
$\delta_{\bigl( [\Lambda_1 , [\Lambda_2, \Lambda_3]^c]^c+\hbox{cyclic}
\bigr) } =0$, and, as a consequence
$(\, [\Lambda_1 , [\Lambda_2, \Lambda_3]^c\,]^c+\hbox{cyclic}\,)$
must be either
zero or a gauge parameter that generates no gauge transformation. The latter
is true, the Jacobi identity for $[\,\, , \, ]^c$, calculated using
\tent\ gives a result of the form $\{ S , \chi \}$, which,
as a gauge parameter,
generates no gauge transformation.

The above observations indicate that $[\,\, , \, ]^c$ leads to a {\it strict
Lie bracket} in the space of even functions
$\Lambda$ modulo functions of the form $\{ S , \chi \}$. Let us denote
the equivalence relation by $\approx$, that is $\Lambda \approx \Lambda
+\{ S , \chi \} $.  Indeed, the
definition given above implies that $[ \Lambda ,
\{ S, \chi \}\, ]^c \approx 0$,
and therefore the bracket depends only on the equivalence class of the gauge
parameters. This bracket defines the Lie algebra of classical gauge
transformations.

\subsection{Quantum gauge transformations.}
The main difficulty in understanding the notion of gauge
transformations at the quantum level was due to the apparent lack of a
suitable invariant object. In Darboux coordinates,
the gauge transformations given
in \one\ do not leave invariant the quantum master action,
nor the measure $d\mu e^{S/\hbar}$, as one could naively hope.
This is easily verified using the result
that the variation of {\it any} measure $d\mu$ under a
canonical transformation generated by $K$ is given by $\wh\delta_K d\mu =
2 d\mu \cdot \Delta_{d\mu} K$ (see [\hatazwiebach], Eqn.(3.26)).
Since we are taking $K=\{ S,\Lambda\}$ with $\Delta_{d\mu}\Lambda =0$
($\Lambda$ is linear in fields),
this result, with the help of eqs.\chmeas, \leibnitz,
and the Jacobi identity, leads immediately
to $\wh\delta_K d\mu e^{S/\hbar}=2 d\mu e^{S/\hbar}\cdot \bigl\{
\Delta_{d\mu}S + {1\over 4\hbar}\{S,S\}, \Lambda\bigr\}$.
If instead of a factor
${1\over 4}$ multiplying the $\{ S,S\}$ term, we would have a ${1\over 2}$,
the master equation would
imply invariance. This means that the measure $d\mu_S = d\mu e^{2S/\hbar}$
introduced earlier is actually invariant under the gauge transformations
of Eqn.\one.

We can now easily generalize the result to arbitrary coordinate systems,
and field dependent gauge transformation parameters.
The variation of the measure
$d\mu_S$ under a canonical transformation generated by $K$ is
given by $\wh\delta_K\, d\mu_S = 2d\mu_S \, \Delta_{d\mu_S} K$, and therefore
the condition of invariance is simply $\Delta_{d\mu_S} K=0$,
i.e. $K$
must be an observable in the full quantum theory. A special class of
solutions, representing trivial observables,
is given by,
$$ K \equiv \hbar\, \Delta_{d\mu_S} \Lambda \,=\, \hbar\, \Delta_{d\mu}
\Lambda + \,\{ S , \Lambda \} ,\eqn\nine$$
where invariance follows due to the nilpotency of $\Delta_{d\mu_S}$.
These gauge transformations close under commutation.
Defining the quantum gauge transformation
$\delta_\Lambda^q \equiv \wh\delta_{\,\hbar\Delta_{d\mu_S} \Lambda}$, we find
$$\bigl[\, \delta_{\Lambda_2}^q \, , \, \delta_{\Lambda_1}^q\, \bigr]
= \delta^q_{[\Lambda_1,\Lambda_2]^q}, \quad\
[\Lambda_1,\Lambda_2]^q \equiv \half \bigl\{ \, \Lambda_1\, ,\, \hbar
\Delta_{d\mu_S}\Lambda_2
 \,\bigr\} -\half\bigl\{ \, \Lambda_2 ,\,\hbar\Delta_{d\mu_S}
 \Lambda_1\,\bigr\}\,.\eqn\tentta$$
The bracket $[\,\, , \,]^q$
differs from its classical counterpart $[\,\, , \,]^c$, given in Eqn.\tent,
by terms of order $\hbar$. In exact analogy to the classical case, we have that
$(\, [\Lambda_1 , [\Lambda_2, \Lambda_3]^q\,]^q+\hbox{cyclic}\,)$
is of the form $\Delta_{d\mu_S} \chi$, which, as a gauge parameter,
generates no gauge transformation.
Therefore $[\,\, , \, ]^q$ leads to a {\it strict
Lie bracket} in the space of even functions
$\Lambda$ modulo functions of the form $\Delta_{d\mu_S} \chi$.
We can check that
$[ \Lambda , \Delta_{d\mu_S} \chi \,]^q \approx 0$,
and therefore the bracket depends only on the equivalence class
$\Lambda \approx \Lambda +\Delta_{d\mu_S} \chi $ of the gauge
parameters. This bracket defines the Lie algebra of quantum gauge
transformations.

\subsection{Example I: Scalar Field Theory}
We first illustrate our ideas with the help of the simplest theory, namely
the theory of a free scalar field $\phi$ in $D$ dimensions described by the
action $S=\int d^Dx (\p_\mu\phi \p^\mu\phi -V(\phi))$. In this case the
classical theory does not possess any gauge invariance
in the usual sense.
The BV formulation
of the theory involves the field $\phi$ and its anti-field $\phi^*$, and
the BV master action coincides with the classical action $S$. As a result,
any local function $K (\phi)$, independent of antifields,
corresponds to a local observable
($\Delta_{d\mu_S}K(\phi) =0$),
and hence generates a gauge symmetry.\foot{Note that a general
$K$ of this form cannot be written as $\Delta_{d\mu_S}\Lambda$ for a
{\it local} $\Lambda$.}
The resulting transformations are
$$ \delta\phi =\{\phi \, , \, K(\phi)\}=0, \quad
\delta\phi^* = \{\phi^* \, , \, K(\phi)\} = -{\delta K(\phi) \over
\delta \phi} \eqn\ephivariation $$
This can easily be seen to be a symmetry of the theory leaving both the
action $S$ (which is independent
of $\phi^*$), and the measure $d\phi d\phi^*$ separately invariant.
This, of course, need not be the case for general $K (\phi,\phi^*) = \hbar
\Delta_{d\mu_S} \Lambda (\phi,\phi^*)$.

\subsection{Example II: Gauge Transformations in Closed String Field Theory.}
The closed string field theory master action is given by [\zwiebachlong]
$$ S = \sum_{g=0}^\infty \hbar^g  \sum_{N=2 \, \for \, g=0\atop
N=1 \, \for \, g\ge 1}^\infty {1\over N!} \, \,
{}_{1\cdots N}\bra{V^{(g,N)}} \Psi\rangle_1
\cdots \ket{\Psi}_N \, . \eqn\esftaction$$
The corresponding measure is $d\mu=\prod_i d\psi^i$
(for notation, see refs.[\zwiebachlong, \senzwiebach, \senzwiebachtwo].)
Let us study gauge transformations generated by observables
of the form $K=\hbar\Delta_{d\mu_S}\Lambda$.
The most general form of $\Lambda$ is given by,
$$ \Lambda = \sum_{g=0}^\infty \hbar^g  \sum_{N=1}^\infty {1\over N!} \,
\, {}_{1\cdots N}\bra{\Lambda^{(g,N)}} \Psi
\rangle_1 \cdots \ket{\Psi}_N \, . \eqn\esftlambda  $$
Separating out the contribution of the terms involving
$\bra{V^{(0,2)}}= \bra{\omega_{12}} Q^{(2)}$, we get,
$$\eqalign{K= & \,\,\hbar\Delta_{d\mu_S} \Lambda  \equiv
 \sum_{g,N\geq 0} \hbar^g \,{1\over N!}
\bra{K^{(g,N)}}\Psi\rangle_1 \cdots \ket{\Psi}_N \cr
= & - \sum_{g\geq 0, N\geq 1}\hskip-6pt\hbar^g \,{1\over N!}\, \,
\bra{\Lambda^{(g,N)}} \sum_{i=1}^N Q^{(i)}
\ket{\Psi}_1 \cdots \ket{\Psi}_N  \cr
& -\sum_{g,N\geq 0}\hskip-6pt\hbar^g\,
\sum_{g_1=0}^g
\sum_{m=1}^{N-1 \, \for \, g_1=g \atop  N+1 \, \for \, g_1<g}
{1\over (N\hskip-2pt -\hskip-2pt m\hskip-2pt +\hskip-2pt 1)!
(m\hskip-2pt -\hskip-2pt 1)!}\,\,{}_{1\ldots }
\bra{V^{(g-g_1, N-m+2)}}\otimes \,
{}_{1'\ldots }
\bra{\Lambda^{(g_1, m)}} \s_{11'}\rangle\cr
& \qquad \qquad  \qquad \qquad \qquad \qquad\quad \cdot\ket{\Psi}_2 \cdots
\ket{\Psi}_{N-m+2} \ket{\Psi}_{2'}\cdots \ket{\Psi}_{m'}\cr &
-{1\over 2} \sum_{g\geq 1,N\geq 0}\hskip-6pt\hbar^g \, {1\over N!}
\, \, \bra{\Lambda^{(g-1, N+2)}}\s_{12}\rangle \ket{\Psi}_3
\cdots \ket{\Psi}_{N+2}\, , \cr} \eqn\edeltasmulambda $$
where $\ket{\s}$
denotes the sewing ket [\senzwiebach].
This generates the gauge transformation:
$$\delta^q_\Lambda\ket{\Psi}_e = \sum_{g,N\geq 0} \hbar^g
{1\over N!} \, \, \bra{K^{(g,N+1)}}\s_{0e}\rangle\ket{\Psi}_1
\cdots \ket{\Psi}_N\, . \eqn\esftgauge$$
In particular, choosing a $\Lambda$ for which only
$\bra{\Lambda^{(0,1)}}\equiv \bra{\Lambda}$ is non-zero, we get
$$\eqalign{
\delta^q_\Lambda\ket{\Psi}_e = & \, - \, {}_0\bra{\Lambda} Q^{(0)}
\ket{\s_{0e}} - \sum_{g=0}^\infty \hbar^g
\hskip-6pt\sum_{N\geq 1\, (g=0)\atop N\geq 0 \,(g>0)}^\infty
{1\over N!} \, \, \bra{V^{(g, N+2)}}  \, {}_0\bra{\Lambda}
\s_{01}\rangle \ket{\s_{2e}}
\ket{\Psi}_{3} \cdots \ket{\Psi}_{N+2}\cr
=\,\, &Q \ket{\Lambda}_e +\sum_{g=0}^\infty \hbar^g
\hskip-6pt\sum_{N\geq 1\, (g=0)\atop N\geq 0 \,(g>0)}^\infty
{1\over N!}\, \,\bra{V^{(g, N+2)}}\s_{1e}\rangle\ket{\Lambda}_2
 \ket{\Psi}_3 \cdots \ket{\Psi}_{N+2}\, , \cr} \eqn\esftspecialgauge$$
where $\ket{\Lambda}_e \equiv \,{}_0\bra{\Lambda}\s_{0e}\rangle$.
Note that the $\ket{\Psi}$ independent term of the gauge
transformation receives contribution from two-punctured surfaces
of all genera.
This means that the notion of an unbroken gauge
symmetry at the classical level differs from the corresponding
notion at the quantum level. If we truncate
to $g=0$ we recover the standard gauge transformations of classical
closed string field theory.

Are there gauge transformations generated by nontrivial observables?
To answer this we need to see
if there are local\foot{Here by local expressions, we
mean terms that when expressed in momentum space, the integrands
have well defined Taylor series expansion in momenta about the point
where all momenta vanish. In position space
these terms will be represented by a series containing higher derivative
terms. In this limited sense the string field theory lagrangian
and the gauge transformations are local, since they contain integration
over subspaces of the moduli space which do not include
any degeneration point.}
observables in string field theory which are not of
the form $\Delta_{d\mu_S}\Lambda$.
It might perhaps be possible to
construct nontrivial local observables  that are quadratic and higher orders
in the fields.
They would generate transformations
$\delta\ket{\Psi}$ that are linear and higher orders in $\ket{\Psi}$, but
do not contain any $\ket{\Psi}$ independent piece. Although these would give
rise to local symmetries of the theory, they would not be
gauge symmetries in the conventional sense of the term.
If we want a symmetry transformation that
contains a field independent piece, we need a $K$ that is linear in the
string field $\ket{\Psi}$. We shall now argue that in string field theory
there is no {\it local} observable that is linear in $\ket{\Psi}$ and
is not of the form $\Delta_{d\mu_S}\Lambda$.
The intuition is simple, such
transformations would correspond at the linearized level to shifts
of the type $\ket{\Psi} \to \ket{\Psi} + \ket{H^Q}$, where $\ket{H^Q}$
is an element of the cohomology of $Q$. Such transformations leave the
kinetic term of the string field action invariant, but certainly do not
qualify as gauge transformations.

To prove this it is enough to take  $K=\bra{K}\Psi\rangle$
linear in $\ket{\Psi}$, and analyze the $\Psi$ independent
terms in the equation $\Delta_{d\mu_S}K=0$. This gives,
$Q^{(e)} \, \,{}_0\bra{K} \s_{0e}\rangle = 0$, and shows
that  ${}_0\bra{K}\s_{0e}\rangle$ must be a BRST invariant
operator. Furthermore, since we want to exclude solutions of the form
$\Delta_{d\mu_S}\Lambda$, we must also require that
${}_0\bra{K}\s_{0e}\rangle$
be a non-trivial member of the BRST cohomology. But in string theory we know
that non-trivial members of the BRST cohomology are found only for
certain specific values of momentum $k^\mu$, satisfying mass-shell
constraints of the form $k^2=m^2$, where $m$ is the mass of one of the
particles in the spectrum of string theory. Thus, when expressed as
a momentum space integral, $\bra{K}\Psi\rangle$ will contain a factor
of $\delta(k^2-m^2)$ in the integrand, and hence it does not correspond
to a {\it local} observable in the theory.

In this example we have found the quantum gauge transformations of
closed string field theory. An obvious question is whether our formalism
can help us understand the gauge structure of string theory. Since the
present approach deals with Lie algebras, it may provide an alternative
or complementary approach to current studies based on homotopy Lie algebras
[\wittenzwiebach].
It remains to be seen if the Lie brackets $[\,\,,\,]^c$ and $[\,\,,\,]^q$
define  managable structures in string
theory. Another interesting project would be to isolate from
the string algebra the Lie subalgebra representing coordinate transformations
in string theory [\ghoshalsengauge].

\singlespace
\refout
\end